\documentclass[aps,prl,preprint,tightenlines,superscriptaddress,showpacs,byrevtex]{revtex4}
\usepackage{graphicx} 
\usepackage{dcolumn}  

\graphicspath{{ps}}

\begin{document}

\vspace*{-3\baselineskip}
\resizebox{!}{3cm}{\includegraphics{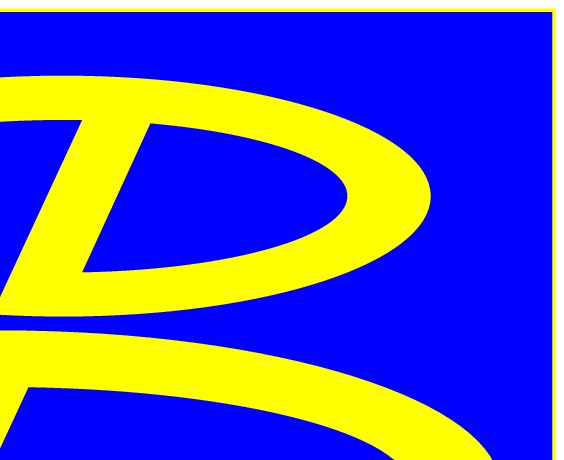}}

\preprint{\vbox{ \hbox{   }
                 \hbox{BELLE-CONF-0426}
                 \hbox{ICHEP04 8-0673}
}}

\title{\boldmath Moments of the Hadronic Mass Spectrum\\
  in Inclusive Semileptonic $B$~Decays at Belle}

\affiliation{Aomori University, Aomori}
\affiliation{Budker Institute of Nuclear Physics, Novosibirsk}
\affiliation{Chiba University, Chiba}
\affiliation{Chonnam National University, Kwangju}
\affiliation{Chuo University, Tokyo}
\affiliation{University of Cincinnati, Cincinnati, Ohio 45221}
\affiliation{University of Frankfurt, Frankfurt}
\affiliation{Gyeongsang National University, Chinju}
\affiliation{University of Hawaii, Honolulu, Hawaii 96822}
\affiliation{High Energy Accelerator Research Organization (KEK), Tsukuba}
\affiliation{Hiroshima Institute of Technology, Hiroshima}
\affiliation{Institute of High Energy Physics, Chinese Academy of Sciences, Beijing}
\affiliation{Institute of High Energy Physics, Vienna}
\affiliation{Institute for Theoretical and Experimental Physics, Moscow}
\affiliation{J. Stefan Institute, Ljubljana}
\affiliation{Kanagawa University, Yokohama}
\affiliation{Korea University, Seoul}
\affiliation{Kyoto University, Kyoto}
\affiliation{Kyungpook National University, Taegu}
\affiliation{Swiss Federal Institute of Technology of Lausanne, EPFL, Lausanne}
\affiliation{University of Ljubljana, Ljubljana}
\affiliation{University of Maribor, Maribor}
\affiliation{University of Melbourne, Victoria}
\affiliation{Nagoya University, Nagoya}
\affiliation{Nara Women's University, Nara}
\affiliation{National Central University, Chung-li}
\affiliation{National Kaohsiung Normal University, Kaohsiung}
\affiliation{National United University, Miao Li}
\affiliation{Department of Physics, National Taiwan University, Taipei}
\affiliation{H. Niewodniczanski Institute of Nuclear Physics, Krakow}
\affiliation{Nihon Dental College, Niigata}
\affiliation{Niigata University, Niigata}
\affiliation{Osaka City University, Osaka}
\affiliation{Osaka University, Osaka}
\affiliation{Panjab University, Chandigarh}
\affiliation{Peking University, Beijing}
\affiliation{Princeton University, Princeton, New Jersey 08545}
\affiliation{RIKEN BNL Research Center, Upton, New York 11973}
\affiliation{Saga University, Saga}
\affiliation{University of Science and Technology of China, Hefei}
\affiliation{Seoul National University, Seoul}
\affiliation{Sungkyunkwan University, Suwon}
\affiliation{University of Sydney, Sydney NSW}
\affiliation{Tata Institute of Fundamental Research, Bombay}
\affiliation{Toho University, Funabashi}
\affiliation{Tohoku Gakuin University, Tagajo}
\affiliation{Tohoku University, Sendai}
\affiliation{Department of Physics, University of Tokyo, Tokyo}
\affiliation{Tokyo Institute of Technology, Tokyo}
\affiliation{Tokyo Metropolitan University, Tokyo}
\affiliation{Tokyo University of Agriculture and Technology, Tokyo}
\affiliation{Toyama National College of Maritime Technology, Toyama}
\affiliation{University of Tsukuba, Tsukuba}
\affiliation{Utkal University, Bhubaneswer}
\affiliation{Virginia Polytechnic Institute and State University, Blacksburg, Virginia 24061}
\affiliation{Yonsei University, Seoul}
  \author{K.~Abe}\affiliation{High Energy Accelerator Research Organization (KEK), Tsukuba} 
  \author{K.~Abe}\affiliation{Tohoku Gakuin University, Tagajo} 
  \author{N.~Abe}\affiliation{Tokyo Institute of Technology, Tokyo} 
  \author{I.~Adachi}\affiliation{High Energy Accelerator Research Organization (KEK), Tsukuba} 
  \author{H.~Aihara}\affiliation{Department of Physics, University of Tokyo, Tokyo} 
  \author{M.~Akatsu}\affiliation{Nagoya University, Nagoya} 
  \author{Y.~Asano}\affiliation{University of Tsukuba, Tsukuba} 
  \author{T.~Aso}\affiliation{Toyama National College of Maritime Technology, Toyama} 
  \author{V.~Aulchenko}\affiliation{Budker Institute of Nuclear Physics, Novosibirsk} 
  \author{T.~Aushev}\affiliation{Institute for Theoretical and Experimental Physics, Moscow} 
% ...
\collaboration{The Belle Collaboration}

\begin{abstract}
  We report a measurement of the first and second moment of the
hadronic mass distribution~$M_X$ in $B\to X_c\ell\nu$~decays for
lepton threshold momenta ranging from 0.9 to 1.6~GeV/$c$ in the
center of mass frame. The measurement uses $B\bar B$ events in which
the hadronic decay of one $B$~meson is fully reconstructed and a
charged lepton from the decay of the other $B$~meson is
identified. These results are obtained from a 140~fb$^{-1}$ data
sample collected near the $\Upsilon(4S)$~resonance with the Belle
detector at the KEKB asymmetric energy $e^+e^-$~collider.

\end{abstract}

\pacs{13.20.He,14.40.Nd}

\maketitle

\tighten

{\renewcommand{\thefootnote}{\fnsymbol{footnote}}}
\setcounter{footnote}{0}

The heavy quark effective theory (HQET) combined with the operator
product expansion provides a framework in which inclusive $B$~decay
properties can be calculated. In particular, the moments of various
inclusive quantities in $B\to X_c\ell\nu$~decays, such as the lepton
momentum~$p_\ell$, the 4-momentum transfer squared~$q^2$ and the
hadronic recoil mass~$M_X$, can be related to the non-perturbative
parameters that also appear in calculations of the semileptonic decay
rate (see Ref.~\cite{ref:1,ref:2,ref:3} for calculations of the
moments of $M_X$). Measurements of these moments can thus improve
calculations of the semileptonic $B$~decay rate and lead to a more
precise determination of the magnitude of the
Cabibbo-Kobayashi-Maskawa (CKM) matrix element
$V_{cb}$~\cite{ref:4}. In this paper, we present a measurement of the
first and second moment of $M_X$, $\langle M_X\rangle$ and $\langle
M^2_X\rangle$, for lepton momentum thresholds ranging from 0.9 to
1.6~GeV/$c$ in the center-of-mass (c.m.) frame.

The analysis is based on the data recorded with the Belle
detector~\cite{ref:5} at the asymmetric energy $e^+e^-$~collider
KEKB~\cite{ref:6}, operating at a c.m.\ energy near the
$\Upsilon(4S)$~resonance. KEKB consists of a low energy ring (LER)
of 3.5 GeV positrons and a high energy ring (HER) of 8 GeV
electrons. The Belle detector is a large-solid-angle magnetic
spectrometer consisting of a three-layer silicon vertex detector
(SVD), a 50-layer central drift chamber (CDC), an array of aerogel
threshold \v{C}erenkov counters (ACC), a barrel-like arrangement of
time-of-flight scintillation counters (TOF), and an electromagnetic
calorimeter comprised of CsI(Tl) crystals (ECL) located inside a
super-conducting solenoid coil that provides a 1.5~T magnetic
field. The responses of the ECL, CDC ($dE/dx$) and ACC detectors are
combined to provide clean electron identification. Muons are
identified in the instrumented iron flux-return (KLM) located outside
of the coil. Charged hadron identification relies on the information
from the CDC, ACC and TOF~sub-detectors.

The $\Upsilon(4S)$~dataset used for this study corresponds
to an integrated luminosity of 140~fb$^{-1}$, or about
152 million $B\bar B$~events. Another 15~fb$^{-1}$ taken 60~MeV below
the resonance are used to subtract the non-$B\bar B$ background. Full
detector simulation based on GEANT~\cite{ref:7} is applied to Monte
Carlo (MC) simulated events. The size of the MC samples is
equivalent to about three times the integrated luminosity. The decay
$B\to D^*\ell\nu$ is simulated using a HQET-based
model~\cite{ref:8}. The ISGW2 model~\cite{ref:9} is used for the
decays $B\to D\ell\nu$ and $B\to D^{**}\ell\nu$. The non-resonant
$B\to D^{(*)}\pi\ell\nu$~component is generated according to the
model of Goity and Roberts~\cite{ref:10}.

We use $B\bar B$~events in which the hadronic decay of one
$B$~meson is fully reconstructed ($B_\mathrm{tag}$) and the
semileptonic decay of the recoiling $B$~meson ($B_\mathrm{signal}$) is
inferred from the presence of a charged lepton (electron or muon). After
selecting hadronic events~\cite{ref:11}, we reconstruct
$B_\mathrm{tag}$ by searching the decay modes $B^+\to\bar
D^{(*)0}\pi^+, \bar D^{(*)0}\rho^+, \bar D^{(*)0}a_1^+$ and $B^0\to
D^{(*)-}\pi^+, D^{(*)-}\rho^+, D^{(*)-}a_1^+$~\cite{ref:12}.
Pairs of photons satisfying $E_\gamma>50$~MeV and
117~MeV/$c^2<m(\gamma\gamma)<150$~MeV/$c^2$ are combined to form
$\pi^0$~candidates. $K^0_\mathrm{S}$~mesons are reconstructed from pairs of
oppositely charged tracks with invariant mass within $\pm
30$~MeV/$c^2$ of the $K^0_\mathrm{S}$~mass and decay vertex displaced
from the interaction point. Candidate $\rho^+$ and $\rho^0$~mesons are
reconstructed in the $\pi^+\pi^0$ and $\pi^+\pi^-$ decay modes,
requiring their invariant masses to be within $\pm 150$~MeV/$c^2$ of the
$\rho$~mass. Candidate $a_1^+$~mesons are obtained by combining a
$\rho^0$~candidate with a charged pion and requiring an invariant mass
between 1.0 and 1.6~GeV/$c^2$.
$D^0$~candidates are searched for in the $K^-\pi^+$, $K^-\pi^+\pi^0$,
$K^-\pi^+\pi^+\pi^-$, $K^0_\mathrm{S}\pi^+\pi^-$ and
$K^0_\mathrm{S}\pi^0$~decay modes. The $K^-\pi^+\pi^+$ and
$K^0_\mathrm{S}\pi^+$~modes are used to reconstruct
$D^+$~mesons. Charmed mesons are selected in a window corresponding to
$\pm 3$ times the mass resolution in the respective decay
mode. $D^{*+}$~mesons are reconstructed by pairing a charmed meson
with a low momentum pion, $D^{*+}\to D^0\pi^+,D^+\pi^0$. The decay
modes~$D^{*0}\to D^0\pi^0$ and $D^{*0}\to D^0\gamma$ are used to
search for neutral charmed vector mesons.

\begin{figure}
  \begin{center}
    \includegraphics{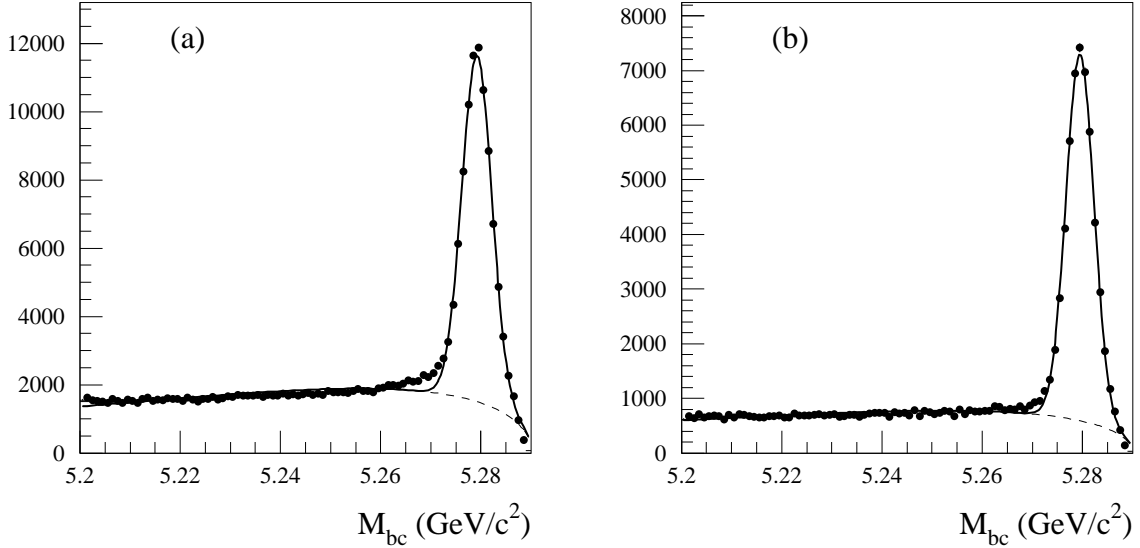}
  \end{center}
  \caption{Reconstruction of the tag-side $B$~meson. The
    $M_\mathrm{bc}$~distribution after requiring $\Delta E<50$~MeV is
    shown for fully reconstructed (a) $B^+$ (b) $B^0$ decays. The fit
    uses a Gaussian function for the signal and an ARGUS background
    function~\cite{ref:14} for the combinatorial background.}
    \label{fig:1}
\end{figure}

For each $B_\mathrm{tag}$~candidate, the beam-constrained
mass~$M_\mathrm{bc}$ and $\Delta E$ are calculated~\cite{ref:13}
\begin{equation}
  M_\mathrm{bc} = \sqrt{(E^*_\mathrm{beam})^2-(\vec p^*_B)^2}~, \quad
  \Delta E = E^*_B-E^*_\mathrm{beam}~,
\end{equation}
where  $E^*_\mathrm{beam}$ is the beam energy in the c.m.\ system,
$\vec p^*_B$ and $E^*_B$ are the 3-momentum and the energy of the
$B_\mathrm{tag}$~candidate, respectively. The signal region for
$B_\mathrm{tag}$ is defined by the selections
$M_\mathrm{bc}>5.27$~GeV/$c^2$ and $\Delta E<50$~MeV. If multiple
candidates are found in a single event, the best candidate is chosen
based on $\Delta E$ and other variables. Fig.~\ref{fig:1} shows the
$M_\mathrm{bc}$ distribution for $\Delta E<50$~MeV. The number of the
fully reconstructed $B^+$ ($B^0$) events, obtained by fitting this
distribution with the sum of a Gaussian function for the signal
component and an ARGUS background function~\cite{ref:14}, is
$76,155\pm 511$ ($46,863\pm 374$) with a purity of the 74\% (81\%).

Semileptonic decays of $B_\mathrm{signal}$ are selected by requiring
exactly one identified charged lepton among the particles that are not
associated to $B_\mathrm{tag}$. In the lepton momentum range relevant
to this analysis, electrons (muons) are selected with an efficiency of
92\% (89\%) and the probability to misidentify a pion as an electron
(a muon) is 0.25\% (1.4\%)~\cite{ref:15,ref:16}. For $B^+$~tags, we
require $Q_\ell\cdot Q_B<0$, where $Q_\ell$ and $Q_B$ are the charges
of the lepton and of $B_\mathrm{tag}$, respectively.

The charged tracks and neutral clusters which are associated neither
with $B_\mathrm{tag}$ nor with the charged lepton are assigned to the
hadronic $X$~system. The 4-momenta of the charged particles are
calculated with the pion mass except for identified kaons. The
missing 4-momentum in the event is calculated,
\begin{equation}
  p_\mathrm{miss} =
  (p_\mathrm{LER}+p_\mathrm{HER})-p_{B_\mathrm{tag}}-p_\ell-p_X~,
\end{equation}
where the indices LER and HER refer to the colliding beams. As only
the neutrino in $B\to X_c\ell\nu$ should be missing in the event, we
require the missing mass to be consistent with zero,
$|M^2_\mathrm{miss}|<2$~GeV$^2$/$c^4$.

The neutrino 4-momentum is inferred from the missing momentum,
$p_\nu=(|\vec p_\mathrm{miss}|,\vec p_\mathrm{miss})$, and the
4-momentum of the $X$~system is recalculated,
\begin{equation}
  p'_X =
  (p_\mathrm{LER}+p_\mathrm{HER})-p_{B_\mathrm{tag}}-p_\ell-p_\nu~.
\end{equation}
The $M_X$~resolution (defined as half width at the half maximum)
obtained with this procedure is about 200~MeV/$c^2$.

\begin{figure}
  \begin{center}
   \includegraphics{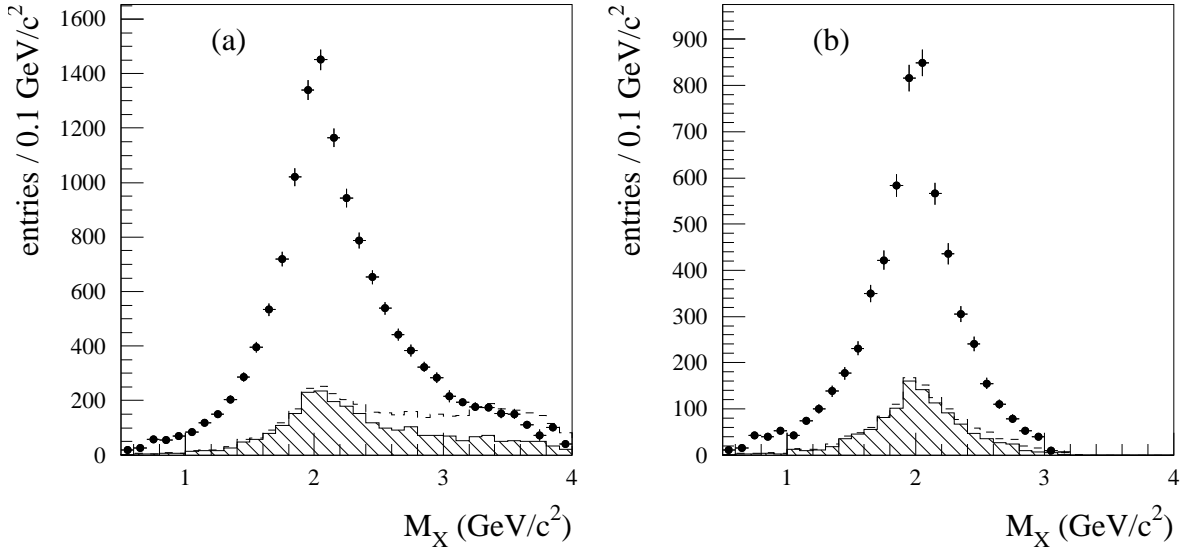}
  \end{center}
  \caption{The $M_X$~distributions for (a) $p^*_\ell>0.9$~GeV/$c$ and
   (b) $p^*_\ell>1.6$~GeV/$c$. The data points are the
   $\Upsilon(4S)$~data after subtraction of the non-$B\bar
   B$~background. The background from secondary or misidentified
   leptons is indicated by the dashed line. The combinatorial
   background in the $B_\mathrm{tag}$~sample is shown as a hatched
   histogram.}
   \label{fig:2}
\end{figure}

The background consists of the following components: combinatorial
background in the $B_\mathrm{tag}$~sample, background from
secondary or misidentified leptons, and background from non-$B\bar
B$~events. The combinatorial background is dominant. Its
shape (in $M_X$ and $M^2_X$) is determined from the MC simulation, and
the $M_\mathrm{bc}$~side-band region
($5.2$~GeV/$c^2<M_\mathrm{bc}<5.27$~GeV/$c^2$, $|\Delta E|<50$~MeV) is used to
determine its normalization, separately for each lepton momentum
threshold. The background from secondary or misidentified leptons is
significant in the low lepton momentum region. Its shape is estimated
from the MC simulation, and the number of well-reconstructed
$B_\mathrm{tag}$ candidates is used for its normalization. The
magnitude of the background from misidentified leptons is corrected
using the pion misidentification rate measured in
$K^0_\mathrm{S}\rightarrow\pi^+\pi^-$~decays in hadronic events. The
background from non-$B\bar B$~events is smallest, and it is subtracted
using the data taken 60~MeV below the
$\Upsilon(4S)$~resonance. Fig.~\ref{fig:2} shows the
$M_X$~distributions for different minimum lepton momenta in the c.m.\
frame, together with the background estimation.

To correct for the distortions caused by finite detector resolution,
the background-subtracted $M_X$ and $M^2_X$~distributions are unfolded
using the Singular Value Decomposition (SVD)
algorithm~\cite{ref:17}. The approach consists in describing the
detector response by a correlation matrix between the true and the
reconstructed $M_X$ ($M^2_X$) values which is determined
from the MC simulation. This matrix is decomposed into its eigenvalue
components, and only the contributions leading to a statistically
stable solution are kept. For the unfolding of the $M_X$ ($M^2_X$)
distribution, the measured spectrum is divided into 35 (39) 0.1
GeV/$c^2$ (0.333 GeV$^2$/$c^4$) wide bins. For the unfolded
distribution, 12 (13) bins are used.

\begin{figure}
  \begin{center}
   \includegraphics{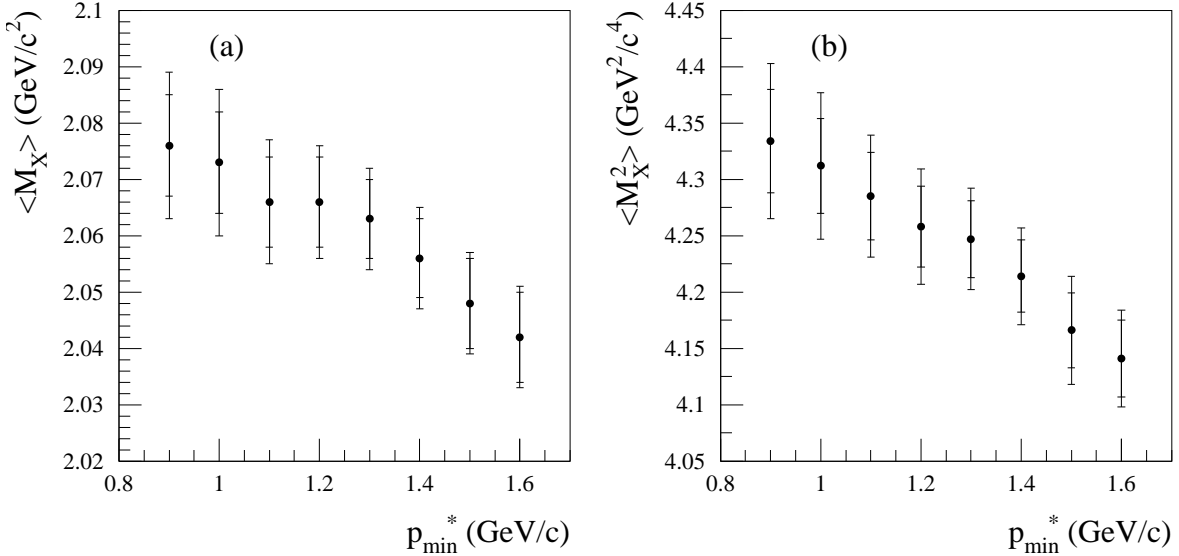}
  \end{center}
  \caption{The first and second hadronic mass moments, $\langle
    M_X\rangle$ and $\langle M^2_X\rangle$, for different lepton threshold
    momenta. The error bars indicate the statistical and total errors.
    Note that the individual moments are highly correlated. All
   results are preliminary.}
  \label{fig:3}
\end{figure}

For each minimum lepton momentum, ranging from 0.9 to 1.6~GeV/$c$, the
background-subtracted $M_X$ and $M_X^2$~distributions are unfolded,
separately for  $B^+$ and $B^0$~tags, and for electron and muon
events. For each of these four sub-samples, the values of $\langle
M_X\rangle$ and $\langle M^2_X\rangle$ are obtained by calculating the
mean values of the unfolded $M_X$ and $M^2_X$~distributions. The
results obtained by averaging the four sub-sample results are shown in
Fig.~\ref{fig:3} and Tables~\ref{tab:1} and \ref{tab:2}. All results
are preliminary. This procedure has been tested on MC simulated events
for the full minimum lepton momentum range and no significant bias has
been found.
\begin{table}
  \begin{center}
    \begin{tabular}{c@{\extracolsep{.3cm}}cccc}
      \hline \hline
      \rule{0pt}{2.7ex}$p^*_{min}$ & $\langle M_X\rangle$ & Detector/
      & Unfolding & $X_c$~model\\
      \rule[-1.3ex]{0pt}{1.3ex}(GeV/$c$) & (GeV/$c^2$) & background &
      &\\
      \hline
      \rule{0pt}{2.7ex}0.9 & $2.076\pm 0.009\pm 0.010$ & 0.006 & 0.008
      & 0.003\\
      1.0 & $2.073\pm 0.009\pm 0.010$ & 0.005 & 0.008 & 0.003\\
      1.1 & $2.066\pm 0.008\pm 0.008$ & 0.005 & 0.005 & 0.003\\
      1.2 & $2.066\pm 0.008\pm 0.006$ & 0.004 & 0.004 & 0.003\\
      1.3 & $2.063\pm 0.007\pm 0.006$ & 0.004 & 0.005 & 0.003\\
      1.4 & $2.056\pm 0.007\pm 0.006$ & 0.003 & 0.004 & 0.003\\
      1.5 & $2.048\pm 0.008\pm 0.005$ & 0.003 & 0.003 & 0.002\\
      \rule[-1.3ex]{0pt}{1.3ex}1.6 & $2.042\pm 0.008\pm 0.004$ & 0.003
      & 0.001 & 0.002\\
      \hline \hline
    \end{tabular}
  \end{center}
  \caption{The first hadronic mass moment $\langle M_X\rangle$ for
    different lepton threshold momenta. The first error on $\langle
    M_X\rangle$ is statistical and the second is total systematic
    uncertainty, {\it i.e.}, the quadratic sum of the right-most three
    columns. All results are preliminary.}
  \label{tab:1}
\end{table}
\begin{table}
  \begin{center}
    \begin{tabular}{c@{\extracolsep{.3cm}}cccc}
      \hline \hline
      \rule{0pt}{2.7ex}$p^*_{min}$ & $\langle M^2_X\rangle$ & Detector/
      & Unfolding & $X_c$~model\\
      \rule[-1.3ex]{0pt}{1.3ex}(GeV/$c$) & (GeV$^2$/$c^4$) &
      background & &\\
      \hline
      \rule{0pt}{2.7ex}0.9 & $4.334\pm 0.046\pm 0.051$ & 0.041 & 0.027
      & 0.014\\
      1.0 & $4.312\pm 0.042\pm 0.049$ & 0.037 & 0.030 & 0.013\\
      1.1 & $4.285\pm 0.039\pm 0.037$ & 0.032 & 0.017 & 0.012\\
      1.2 & $4.258\pm 0.036\pm 0.036$ & 0.027 & 0.021 & 0.012\\
      1.3 & $4.247\pm 0.034\pm 0.030$ & 0.023 & 0.016 & 0.011\\
      1.4 & $4.214\pm 0.032\pm 0.028$ & 0.024 & 0.009 & 0.011\\
      1.5 & $4.166\pm 0.033\pm 0.035$ & 0.022 & 0.025 & 0.011\\
      \rule[-1.3ex]{0pt}{1.3ex}1.6 & $4.141\pm 0.034\pm 0.027$ & 0.022
      & 0.010 & 0.013\\
      \hline \hline
    \end{tabular}
  \end{center}
  \caption{The second hadronic mass moment $\langle M^2_X\rangle$ for
    different lepton threshold momenta. The first error on $\langle
    M^2_X\rangle$ is statistical and the second is total systematic
    uncertainty, {\it i.e.}, the quadratic sum of the right-most three
    columns. All results are preliminary.}
  \label{tab:2}
\end{table}

We consider three sources of systematic error, shown separately in
columns three to five of Tables~\ref{tab:1} and \ref{tab:2}:
uncertainty related to detector modeling and residual background
subtraction, uncertainty related to the unfolding procedure, and
uncertainty related to the $X_c$~model in the MC
simulation. The total systematic error quoted on the moments
corresponds to the quadratic sum of these three components.

The uncertainty related to the detector modeling and to the background
subtraction is estimated from the consistency between the four
sub-samples, by varying the selections used in the analysis (namely the
$B_\mathrm{tag}$~signal region and the requirement on
$M^2_\mathrm{miss}$), and by varying the background normalization. The
uncertainty related to the unfolding procedure is estimated by
changing the number of eigenvalue components in the unfolding
solution. The $X_c$~model uncertainty  is obtained by varying the
fractions of $B\to D^*\ell\nu$, $B\to D\ell\nu$ and $B\to
D^{**}/D^{(*)}\pi\ell\nu$ within $\pm 10\%$, $\pm 10\%$ and $\pm
30\%$, respectively, and summing the individual variations in
quadrature. These ranges of variation roughly correspond to the
experimental uncertainties~\cite{ref:18} in the isospin averaged
branching ratios.

In summary, we have performed a measurement of the first and second
moment of the hadronic mass distribution $M_X$ in $B\to
X_c\ell\nu$~decays for minimum lepton momenta ranging from 0.9 to
1.6~GeV/$c$ in the c.m.\ frame. The results listed in
Tables~\ref{tab:1} and \ref{tab:2} are compatible with recent values
from other experiments~\cite{ref:20,ref:21}. It is expected that this
measurement (combined with measurements of the semileptonic branching
fraction and of the moments of the lepton spectrum in $B\to
X_c\ell\nu$) will lead to an improved determination of the CKM matrix
element $|V_{cb}|$.

We thank the KEKB group for the excellent operation of the
accelerator, the KEK Cryogenics group for the efficient
operation of the solenoid, and the KEK computer group and
the National Institute of Informatics for valuable computing
and Super-SINET network support. We acknowledge support from
the Ministry of Education, Culture, Sports, Science, and
Technology of Japan and the Japan Society for the Promotion
of Science; the Australian Research Council and the
Australian Department of Education, Science and Training;
the National Science Foundation of China under contract
No.~10175071; the Department of Science and Technology of
India; the BK21 program of the Ministry of Education of
Korea and the CHEP SRC program of the Korea Science and
Engineering Foundation; the Polish State Committee for
Scientific Research under contract No.~2P03B 01324; the
Ministry of Science and Technology of the Russian
Federation; the Ministry of Education, Science and Sport of
the Republic of Slovenia; the National Science Council and
the Ministry of Education of Taiwan; and the U.S.\
Department of Energy.



\begin{thebibliography}{99}
\bibitem{ref:1} P.\ Gambino and N.\ Uraltsev, Eur.\ Phys.\ J.\ {\bf C
  34}, 181 (2004).
\bibitem{ref:2} A.\ Falk and M.\ Luke, Phys.\ Rev. {\bf D 57}, 424
  (1998).
\bibitem{ref:3} A.\ Falk, M.\ Luke and M.\ Savage, Phys.\ Rev.\ {\bf D
  53}, 2491 (1996); {\it ibid}, {\bf 53}, 6316 (1996).
\bibitem{ref:4} M.\ Kobayashi and T.\ Maskawa, Prog.\ Theor.\ Phys.\
  {\bf 49}, 652 (1973).
\bibitem{ref:5} A.\ Abashian {\it et al.} (Belle Collaboration), Nucl.\
  Instr.\ and Meth.\ {\bf A 479}, 117 (2002).
\bibitem{ref:6} S.\ Kurokawa and E.\ Kikutani, Nucl.\ Instr.\ and
  Meth.\ {\bf A 499}, 1 (2003), and other papers included in this
  volume.
\bibitem{ref:7} R.\ Brun {\it et al.}, GEANT 3.21, CERN Report
  DD/EE/84-1 (1984).
\bibitem{ref:8} J.\ Duboscq {\it et al.} (CLEO Collaboration), Phys.\ Rev.\
  Lett.\ {\bf 76}, 3898 (1996).
\bibitem{ref:9} N.\ Isgur and D.\ Scora, Phys.\ Rev.\ {\bf D 52}, 2783
  (1995). See also N.\ Isgur {\it et al.}, Phys.\ Rev.\ {\bf D 39}, 799 (1989).
\bibitem{ref:10} J.L.\ Goity and W.\ Roberts, Phys.\ Rev. {\bf D 51},
  3459 (1995).
\bibitem{ref:11} The selection of hadronic events is described in K.\
  Abe {\it et al.} (Belle Collaboration), Phys.\ Rev.\ {\bf D 64}, 072001
  (2001).
\bibitem{ref:12} Throughout this paper, the inclusion of the charge
  conjugate mode is implied.
\bibitem{ref:13} Throughout this paper, quantities calculated in the
  c.m.\ frame are denoted by an asterisk.
\bibitem{ref:14} H.\ Albrecht {\it et al.} (ARGUS Collaboration), Phys.\
  Lett.\ {\bf B 241}, 278 (1990).
\bibitem{ref:15} K.\ Hanagaki {\it et al.}, Nucl.\ Instr.\ and Meth.\
{\bf A 485}, 490 (2002).
\bibitem{ref:16} A.\ Abashian {\it et al.}, Nucl.\ Instr.\ and Meth.\
{\bf A 491}, 69 (2002).
\bibitem{ref:17} A.\ H\"ocker and V.\ Kartvelishvili, Nucl.\ Instr.\
  Meth.\ {\bf A372}, 469 (1996).
\bibitem{ref:18} S.\ Eidelman {\it et al.}, Phys.\ Lett.\ {\bf B592},
  1 (2004).
\bibitem{ref:19} M.\ Battaglia {\it et al.} (DELPHI Collaboration), DELPHI
  2003-028 CONF 648.
\bibitem{ref:20} S.E.\ Csorna {\it et al.} (CLEO Collaboration), Phys.\
Rev.\ {\bf D 70}, 032002 (2004).
\bibitem{ref:21} B.\ Aubert {\it et al.} (BABAR Collaboration), Phys.\
Rev.\ {\bf D 69}, 111103 (2004).
\end{thebibliography}
\end{document}